# Using almetrics for contextualised mapping of societal impact: from hits to networks

Nicolas Robinson-Garcia[1], Thed N. van Leeuwen[2] and Ismael Ràfols[1,2]

[1]INGENIO (CSIC-UPV), Universitat Politècnica de València, València, Spain
[2]Centre for Science and Technology Studies (CWTS), Leiden University, Leiden, The Netherlands

**Abstract**
In this article, we develop a method that uses altmetric data to analyse researchers' interactions, as a way of mapping the contexts of potential societal impact. In the face of an increasing policy demand for quantitative methodologies to assess societal impact, social media data (altmetrics) has been presented as a potential method to capture broader forms of impact. However, current altmetric indicators were extrapolated from traditional citation approaches and are seen as problematic for assessing societal impact. In contrast, established qualitative methodologies for societal impact assessment are based on interaction approaches. These argue that assessment should focus on mapping the contexts in which engagement among researchers and stakeholders take place, as a means to understand the pathways to societal impact. Following these approaches, we propose to shift the use of altmetric data towards network analysis of researchers and stakeholders. We carry out two case studies, analysing researchers' networks with Twitter data. The comparison illustrates the potential of Twitter networks to capture disparate degrees of policy engagement. We propose that this mapping method can be used as an input within broader methodologies in case studies of societal impact assessment

**Keywords** Societal impact, social engagement, open science, altmetrics, twitter, research evaluation

# 1. Introduction

Despite increasing policy demands to assess the societal impact of research, there are not yet well-established quantitative methodologies or set of indicators developed for this task. The approaches so far developed to capture societal impact are mainly based on qualitative approaches (see Joly et al. 2015), while conventional quantitative indicators are based either on publication and citations or socioeconomic measures that cannot capture the social contribution of research. Thus, policy makers are seeking alternative assessment methods.

As publicly funded institutions, universities and public research organisations are shaped by societal demands and challenges. Since the 1980s, they have been increasingly subjected to external pressures which have affected their governance (Geuna & Muscio 2009 p. 94). Among others, the massification of higher education, the increased scale of research and the globalisation of the higher education landscape have strongly shaped universities and research policies (Hazelkorn 2011). Business-like and New Public Management practices have been incorporated to academia, introducing quantitative measures that aim to offer an 'objective and transparent' view on the performance of scientific organizations. These practices, which aim to improve research management and strategic planning have been largely based on publication and citation analysis focused on measuring scientific (rather than societal) impact (van Leeuwen & Moed 2012).

As a result of these transformations in the university, research evaluation has also changed. Evaluation schemes have shifted from purely scientific criteria, towards perspectives that emphasise both scientific and societal relevance. The rationale behind this shift is that research should respond and contribute to societal challenges and demands (Wallace & Rafols 2015 p. 90). Universities are not only expected to contribute to the creation of knowledge and the training of citizens, but also to engage on social outreach and public engagement. Therefore, assessment methods and indicators should adapt to these expectations.

However, the methodologies and indicators so far developed have been found insufficient to satisfy research policy demands (Wilsdon & al. 2015). This is partly due to multiple understandings associated with what is now called 'societal impact', often referred as 'broader impact' in the US (Frodeman & Parker 2009; Holbrook 2012). Spaapen and van Drooge (2011, p.212) define societal impacts of knowledge as "behavioural changes that happen because of this knowledge". These changes may regard human well-being ('quality of life') and/or the social relations between people or organizations. The UK Research Excellence Framework (REF) describes societal impact as "an effect on, change or benefit to the economy, society, culture, public policy or services, health, the environment or quality of life, beyond academia" (REF, 2011). Societal impact encompasses what was earlier called the 'Third Mission' of academia (Molas-Gallart & Castro-Martínez 2007) or also 'research extension' in the context of agriculture and/or development (Ison & Russell 2007).

The policy discourse on societal impact includes related concepts, in particular a more interactive understanding of the relation between researchers and stakeholders, according to which science and society co-evolve through reciprocal feedbacks (Nowotny et al. 2001). Recent policy developments in various countries illustrate this more interactive notion of the relationship. The assessment of societal impact in the REF in the United Kingdom is the most well-known case (Samuel & Derrick 2015), but it is not the only one. The 'Crowdsourcing and Citizen Science Act of 2015' in the US legislation introduces the concept of a 'Citizen Science', not only understood as the involvement of the citizen in well-defined research projects (Bonney et al. 2009), but as actors participating in the definition of the research agenda, by 'furthering science diplomacy through the worldwide collaboration between scientists and citizens'(*Crowdsourcing and Citizen Science Act of 2015* 2015). More recently, the National Dutch Research Agenda, reflects a collaborative effort involving scientists, citizens and companies for setting the research priorities of the Netherlands for 2015-2020.[1] These changes on how research agendas are designed and assessed also reflect transformations in the governance of science. This can be seen, for example, in the emergence of concepts such as Responsible Research and Innovation (Owen et al. 2012) or Open Science (Schroeder 2007) in the policy discourse.

An area of promise for tracing 'broader' forms of impact is altmetrics, i.e. the use of social media data for tracking research impact (Priem et al. 2012). Although still in their infancy, altmetric indicators are based on the idea that scholarly communication is shifting towards the online environments of social media. Most altmetric analyses have extrapolated the bibliometric citation model to social media, and developed indicators based on the mentions (saving or citations) of scientific publications in social media platforms such as Twitter, Mendeley or blogs. Here we will use 'altmetrics' and 'social media data' as synonyms. One should distinguish altmetric data from 'altmetric measures' or 'altmetric indicators' which are the figures of counts of mentions, savings or citations presented in websites such as Altmetric.com or ImpactStory.org

Since social media platforms are becoming an increasingly important channel of informal communication within the research community and between academia and the public, it has been proposed that measuring social media citations to publications is a proxy of societal impact (e.g., Bornmann 2014). However, altmetrics capability to capture traces of societal impact has also been seriously contested. According to Sugimoto and colleagues (2016, p. 23) "media has rather opened a new channel for informal discussions among researchers, rather than a bridge between the research community and society at large". According to Haustein (2015, p. 6):

> "The availability of big data and the ease with which they can be assessed was met by a growing demand, particularly by research funders and managers, to make the societal

[1] http://www.wetenschapsagenda.nl/?lang=en

> impact of science measurable (…)—despite the current lack of evidence that social media events can serve as appropriate indicators of societal impact. Along these lines, the role of publishers in the development of altmetrics needs to be emphasized. Owned by large for-profit publishers, Altmetric, Plum Analytics and Mendeley operate under a certain pressure to highlight the value of altmetrics and to make them profitable. Similarly, many journals have started to implement altmetrics, not least as a marketing instrument."

This article explores the potential use of altmetrics data for the assessment of societal impact in the light of the concepts and approaches developed for impact assessment in research evaluation. We build on approaches such as SIAMPI (Spaapen and van Drooge 2011) and ASIRPA (Joly et al. 2015), which consider that research evaluation should mainly aim at learning rather than accountability, and should focus on identifying the processes that may lead to societal impact. In this paper, we focus, specifically, in the interactions taking place between academics and non-academics.

We suggest that the potential of social media data for impact assessment lies in that it may help understand the social interactions of researchers, both within and outside the scientific sphere. This implies a shift from the focus on the quantified 'impact' of current altmetrics indicators towards the use of social media networks to reveal the diverse contexts in which researchers potentially participate. As an example of this approach, this paper uses Twitter, a public platform where individuals communicate, often combining private and professional interests and engaging on local and global topics.

We have structured the paper as follows. Section 2 reviews first, the insights from qualitative approaches to societal impact assessment and second, the policy expectations of current altmetric indicators as a potential means to measuring broader forms of impact. Section 3 presents our framework for the analysis of social media data to inform societal impact assessment approaches, in particular those based on case studies and focused on researchers' capability to productively engage with other stakeholders. In section 4 we propose a specific method, based on network analyses of social media, for the identification of the potential contexts of engagement with stakeholders. We present two examples of network analyses of researchers and their social contexts based on Twitter data. These case studies illustrate the potential of social media data to capture the type of networks that researchers inhabit and show how they communicate with their scientific peers and non-academic stakeholders. Finally, section 6 concludes by tentatively proposing a research agenda on the mapping of researchers' engagement (rather than impact) using social media in case studies.

# 2. Societal impact assessment and altmetrics indicators

## 2.1 Approaches to societal impact assessment

Societal impact assessment is one of the key challenges in research evaluation (Spaapen & van Drooge 2011). It is worth noting that fashionable policy concepts or buzzwords such as 'open science and innovation' (Moedas 2015) or 'responsible research and innovation' (Meijer et al. 2016; Owen et al. 2012), also closely relate to the idea that science should contribute to improving society. Therefore, their translation into policy and assessment are often associated to initiatives for assessing societal impact.

The large variety of societal benefits and the multiplicity of ways in which these benefits can be achieved questions the assumption that a single methodology may capture societal impact. Societal impact is presented in the literature as a 'basket concept' which includes socioeconomic, environmental, political or educational impact (Moed & Halevi 2015; Reale et al. 2014). Narrow and unidimensional understandings of societal impact neglect the societal contributions gained from academics' activities in many scientific fields. Since the forms of assessment should be aligned with the type of impacts produced, the diversity in both impacts and pathways to impact explains the proliferation of qualitative and quantitative methodologies (e.g. Molas-Gallart et al. 2002; de Jong et al. 2011; Donovan & Butler 2007).

Yet, in spite of the variety of approaches, recent reviews on the assessment of societal impact highlight a significant consensus regarding how societal impact occurs.[2] Here we build mainly on insights the research evaluation community, in particular the SIAMPI[3] (Spaapen and van Drooge 2011) and ASIRPA projects[4] (Joly et al. 2015), but the understandings underlying these approaches are consistent across different evaluation traditions, such as *Contribution Mapping* from health evaluation (Kok & Schuit 2012), or *PIPA* from evaluation in developing countries (Douthwaite et al., 2007). We call this family of methodologies 'interaction approaches' to societal impact assessment (another name could be 'process-oriented' approaches, but in this article we aim to emphasize the interactional part). Hence, the insights described in the ensuing paragraphs represent a perspective of research impact shared by disparate evaluation communities in science policy, health, agriculture and development.

The first relevant observation, consistently reported by many studies, is that it usually takes a long time from the beginning of the research and the societal impact. This time span is estimated in the order of 15 to 20 years. Since this is too long for evaluative purposes, societal impact evaluation cannot be based on assessment of the final impact.

Second, theories of innovation emphasise that research contributes to innovation via social networks in which effects are not linear and causality cannot be attributed to single factors, but to complex interactions in networks (Freeman, 1991; Callon, 1986). The fact that innovation is networked or co-created means that societal impact cannot be attributed only to research efforts. Instead, research is understood as one among many contributions to impact (and moreover the benefits of impact can be contested). Co-creation and ambiguity in attribution (and value of impact) mean that the assessment should focus on the *processes* of interaction or engagement leading to impact rather than the outputs or the impacts themselves.

Third, given the large variety of ways in which researchers and stakeholders interact to generate societal impact, it is not possible to have universal indicators of impact. Indicators of societal impacts are context-dependent since societal impact results from different processes dependent on specific epistemic, social and local contexts. Therefore, approaches to the assessment of impact are based on case studies rather than based on universal methods.

Fourth and finally, from the previous insights it follows that societal impact assessment cannot and should not mimic conventional citation analyses of scientific impact evaluation. These indicators of scientific performance are mainly used for accountability purposes and are conventionally based on linear and universalistic notions of impact. They rely on full attribution of output and are calculated over a relatively shorts periods of 3 to 5 years. Given the long-time span for societal impact, its networked natured, and the ambiguities in attribution, experts like Spaapen (2015, p. 37) advocate that the prime goal of evaluation should shift 'from accountability to communication between partners [i.e. stakeholders] –regarding goals and research designs— and to learning.'

[2] This consensus does not include altmetric studies, which are often atheoretical (i.e. do not make explicit their social impact model) (see Haustein et al. 2015 for an exception), and generally appear to understand 'impact' as having effects on communication networks rather than broader social practices.
[3] The 'Social Impact Assessment Methods for research funding instruments through the study of Productive Interactions' an EU framework programme project known as SIAMPI (http://siampi.eu).
[4] ASIRPA (Analyse des Impacts de la Recherche Publique Agronomique) is an approach to the assessment of socioeconomic impacts of research developed by the French public research institute of agricultural research, INRA (https://www6.inra.fr/asirpa_eng/)

The limitations of case study approaches are its reliance on self-reporting (which may distort facts) and on evaluators' expertise (which is time consuming and has relatively high economic costs) (Spaapen & van Drooge, 2011). In the face of these challenges, funding agencies continue asking for generalizable methods and indicators which fit with accountability logics and that can reduce the economic and time burdens. To these demands, social media and altmetrics analyses are presented as a promising candidate to capture impact signals. This is reflected, for example in recent policy documents such as the Metric Tide Report (Wilsdon et al. 2015) or the European Commission Expert Group on Altmetrics which will, among other tasks, "define the features of a 'responsible metrics' aimed at a responsible use of altmetrics to advance open science, able to track desirable impacts, and qualities of scientific research".[5] The next section will review the main literature analysing their development as alternative or complementary indicators for research evaluation.

## 2.2 Altmetrics as indicators for research evaluation

Altmetrics are defined as "the creation and study of new metrics based on the Social Web for analysing, and informing scholarship" (Priem et al. 2010). They are characterised by their heterogeneity and their presumed capacity to capture traces of the social activity of researchers.

The concept of 'altmetrics' dates from 2010 when it was coined in a tweet by Priem (2010) and then promoted through the 'altmetrics manifesto' (Priem et al. 2010). Altmetrics can be seen as consistent with the 'open science' movement although they have been controversial due to their ambiguity and the pre-existence of overlapping concepts (Ronald & Fred 2013; Thelwall 2012). Moed (2015) identifies three major drivers on the ascent of altmetrics. First, the demand for a research evaluation framework which conceives science as an activity with multiple dimensions and impacts, in particular the increasing interest on measures for assessing societal impact. Second, the technological transformations in science communication in recent years, which are interpreted as a second wave on the digitisation of scholarly communication (Sugimoto et al. 2016). Third, the 'open science' movement which calls for more accessible and transparent research practices – including the dissemination of research findings with research data and methods (Moed, 2015).

The main assumption of altmetrics is that researchers are using social media platforms to disseminate and discuss scholarly contributions, and that in doing so, they are opening scholarly conversations to broader audiences. Based on such premises, it is assumed that by capturing the discussion of scientific papers in social media, altmetrics can provide measures of societal engagement (Piwowar 2013; Bornmann, 2014). These claims are highly disputed due to both conceptual and technical reasons. Technical issues include discrepancies in results depending on data sources (Zahedi, Fenner & Costas, 2014), difficulties of linking social media discussions with scientific papers (Robinson-García et al. 2014), the heterogeneity of social media platforms used, and the easy gaming of the metrics (Robinson-Garcia et al., forthcoming; Sugimoto et al. 2016). Conceptually, social media platforms are built on the premise that the more they are used, the better, since their aim is to facilitate sharing and linking to contents in an automatic way. The analogy between citing a paper in a scholarly work and citing it in social media does not work in the sense, that there is a clear and conscientious motivation for the former, while the latter may simply be a mechanistic and unreflective act (Robinson-Garcia et al., forthcoming).

One of the critical questions regarding altmetrics' use for evaluation, is the extent to which researchers use social media for communication with their scientific peers and societal stakeholders. Due to its rapid development, usage studies become outdated quickly and findings are often contradictory. Some social media platforms which at one point became popular, disappear later (e.g., Connotea) or others which at first had a prominent role, then apparently ended up with a secondary role (e.g., blogs). In the initial years of altmetrics, studies on scholarly usage of social media reported a low participation of scholars (Cabezas-Clavijo & Torres-Salinas 2010; Tenopir et al. 2013). Its subsequent expansion, especially reflected on the usage of platforms such as Facebook and Twitter, was received at first with scepticism as to their credibility and efficacy as dissemination tools (Grande et al. 2014). However, the coverage of researchers' in social media keeps improving with the raise of academic platforms such as Mendeley, ResearchGate or FigShare, together with the continuing increase on the use of general social media platforms such as Twitter.

[5] To learn more about the tasks and members of this expert group, see http://ec.europa.eu/research/openscience/index.cfm?pg=altmetrics_eg

Another key issue in social media analyses is the heterogeneity of sources and diversity of uses covered under the term of 'altmetrics' (Robinson-García et al. 2014; Sugimoto et al. 2016). As a consequence of this variety of platforms and usages, the coverage of science by altmetrics is extremely uneven across fields (Costas et al. 2015), ages and ranks (Procter et al. 2010), or countries (Alperin 2015a) --being extremely low in many cases. This variability is compounded by the fact that researchers use social media for academic and personal purposes, in a context where academic and individual interests cannot be clearly separated from each other.

Conceptually, there is still a serious lack of understanding on what it means for a publication to be mentioned in the various social media. To our knowledge, only Haustein and colleagues (2015) have recently made efforts to theorize the meaning of different altmetric 'acts'. Haustein and colleagues develop their framework from theories of citations (normative and social) and provide interpretations regarding the meaning of social media 'acts' such as saving in Mendeley, mentioning in Twitter, and citing in a blog post -- interestingly discussing different degrees of engagement in the various 'acts'. However, they do not conclude with an interpretation of the altmetric measures derived from these mentions or citations –whether a proxy for societal impact, early scientific attention, educational use, popularity or visibility. They argue that 'the new metrics are all of the above [impact, use, attention, popularity, etc.] and the extent to which each of these occurs depends on the particular platform, its uptake and users, as well as on the research topic, the unit of analysis, and the context of the metric.' (Haustein et al. 2015, p. 397). This suggests that altmetric measures are problematic as indicators (what do they indicate?) – given that the meaning of the measures is unknown, unclear or ambiguous at best.

Altmetric measures are also problematic as indicators of societal impact given the theoretical understandings of societal impact assessment reviewed in the previous subsection. Speaking very generally, current altmetric measures as presented in Altmetric.com or ImpactStory.org webs are linear (from paper to tweet or blog), universal (counts without much context), based on a full attribution model, on outputs (mentions to papers) and seem to be more aimed at accountability purposes (justification of investments) than at learning. Even the focus on immediacy is questionable given that societal impacts typically occur in the long term. These characteristics of the measures run against the notion that societal impact assessment should be about contribution rather than attribution, should aim at understanding engagement and contexts, and based on case studies – because it is ultimately about fostering learning and improvement.

However, there are indeed more instances of contextualisation in altmetric websites than in conventional bibliometric information, for example the geo-localisation of tweets in Altmetric.com. And whilst altmetric measures (e.g. 58 tweets or 10 blog citations) are of questionable usefulness without further contextualisation, the altmetric website does provide extremely interesting information of each specific instance of mentions in social media which would allow to provide much richer insights. Therefore, we should be careful not throw the baby out with the bathwater.

Social media, and particularly microblogging such as Twitter, have an important strength which seems to be overlooked by some altmetric studies: their potential to partly reflect the different audiences with which researchers engage, thus facilitating the identification of disparate stakeholder communities (Hsu & Park 2011). Some researchers have tapped into this treasure trove of information. For instance, Ke and colleagues (2017) proposed a method for systematically identifying researchers and their field of specialisation in Twitter based on lists and the biodata users include in their profile. However, the method is highly restrictive favouring precision over recall and biased towards elite and highly active science communicators. Another line of enquiry is to understand how the social communities reflected in social media vary depending on the platform used (i.e., Twitter, LinkedIn, etc.) and their capability to reflect the 'real' social communities of scientists. Another means to do so is that developed by (Alperin 2015b) where a bot is used in Twitter which surveys users enquiring on any particular topic.

Hence, social media in general, and specifically platforms such as Twitter, could further our understanding on researchers' communities beyond the scientific realm. However, could social media data be used to inform the assessment of societal impact, as carried out in case studies such as the UK's REF? In the following sections, we present an attempt to use social media (altmetrics) data in such a way that it helps inform about researchers' networks and engagement, providing helpful information for the assessment of their potential societal impact.

# 3. From hits to networks: using altmetric data to capture interactions

The literature review in the previous section has revealed a misalignment between the understandings of societal impact in the evaluation communities and current altmetric indicators. According to evaluation experts, societal impact assessment should try to map the processes and contexts of interaction between researchers and societal stakeholders, whereas current altmetrics focus on the linear counting of traces (citations, mentions, savings) towards potential impact.

Following Spaapen and colleagues, we propose that, instead of emulating the indicators used for scientific impact --based on outputs or citation approaches--, societal impact assessment should develop a different methodological approach --based on interaction approaches (Spaapen, 2015, p. 36; Molas-Gallart et al., 2015). Table 1, illustrates the conceptual differences between what we call the citation versus the interaction approaches. Since current altmetric indicators were extrapolated from citation approaches, they are very problematic for societal impact assessment. We suggest that altmetric data can be usefully employed to illuminate societal impact – but this implies a shift in the forms of analysis towards the interaction approaches described above.[6]

Costas and colleagues (2016) have recently proposed a related distinction by confronting current altmetric indicators (what they call 'evaluative altmetrics') to altmetric analyses that describe contents and contexts (what they call 'descriptive altmetrics). Costas' proposal is inspired by the traditional distinction between descriptive scientometrics (focused on trends, science maps and collaboration networks) versus evaluative scientometrics (focused on quantitative indicators for benchmarking and accountability). As an example of their new and descriptive altmetrics, Costas and colleagues (2016) show how Twitter mentions of scientific papers can inform analysts about the different health interests between geographical regions –AIDS in Africa in constrast to mental health and obesity in Europe.

In this section, we argue why and show how the use of the altmetric data for societal impact assessment should based on interaction approaches. Table 1 summarises the differences between current altmetric literature and the approach we propose here. First, given altmetric data is very incomplete and easy to manipulate, its use for evaluation should focus on improvement and learning; a focus on accountability might also result in gaming. Second, given that the large variety of pathways to impact, the unevenness in coverage and the diversity of platforms (Twitter,

[6] Let us notice that citation approaches are also problematic to assess scientific impact, particularly outside of the natural sciences. Interaction approaches could also be useful in the assessment of scientific impact, but this discussion is beyond the scope of this article.

Facebook, blogs, etc.), altmetrics data need to be interpreted according to context -- therefore analyses should be contextual and feed only into case studies. Third, social media data can be useful to trace communications between stakeholders and therefore reveal processes among stakeholder potentially conducive to societal impact. The analysis of these processes may help understand the type of engagement and contribution of researchers in the community that developed an innovation or brought about social change. An analysis in terms of networks (which can be more or less formal), can facilitate the understanding of the contexts (attributes of nodes), processes (links) and embedding (networks structure) of researchers.

**Table 1.** Differences between citation and interaction approaches to societal impact assessment.

| | **Citation approaches** (most current altmetrics) | **Interaction approaches** (proposed approach to altmetrics) |
|---|---|---|
| **Purpose of evaluation** | Accountability | Learning and improvement |
| **Scope of application** | Universal, generalizable | Contextual, case studies |
| **Focus of analysis** | Outputs | Processes, Interactions |
| **Interaction with stakeholders** | Indirect influence | Engagement |
| **Causality** | Attribution | Contribution |
| **Operationalisation** | Linear counts | Network structure |

## 3.1 Altmetrics analysis as case studies for contextualisation and learning

The initial expectations regarding altmetrics were that their indicators could become a general or 'scalable' method to capture traces of societal impact in a way that was useful for accountability purposes. The empirical findings reviewed above suggest that comparisons of social media presence of researchers or their outputs are dangerous and that universal and generalizable application of altmetrics for evaluation is problematic (Wilsdon, 2015).

Social media data can be still extremely informative and helpful, but the heterogeneity of altmetric data suggests to adopt a case study approach, in which complementary information allows to interpret the meaning of the altmetric traces. Given that methodologies such as SIAMPI or ASIRPA are based on case studies, altmetric data could be garnered to provide supportive evidence for case studies within these broader approaches, aimed at learning about the processes leading to societal impact. In short, the rich variety but serious lack of robustness of altmetric data mean that it can be useful for enriching and triangulating evidence in larger case studies about how societal impact occurs.

## 3.2 Focus on interactions, engagement and contribution

Another change of analytical perspective for societal impact assessment concerns to shift the focus from outputs to processes. Given the long-time span between research and impact, research assessment cannot be based on final impacts but on processes potentially leading to impact. One interesting characteristic of social media data is that they can provide early information about communication patterns between researchers and stakeholders – thus allowing to trace interactions that are potentially 'productive' (Spaapen & van Drooge 2011). From this perspective, outputs such as publications, tweets or blogs are interesting as instances of processes of interaction –and the analytical focus should also look at the interactions rather than the outputs. Such interactions may inform about how researchers engage with other communities, in particular when there is a reciprocal relationship (Molas-Gallart et al. 2015). However, the interpretation of these communication processes has to take into account that researchers are only one of the influences of stakeholders. Mentions in social media should not be read as attribution, since research is only one of the many contributions to societal impact (Spaapen & van Drooge 2011).

### *3.3 Networks description*

Network analysis in the form of science maps has become an increasingly popular tool for conveying results of descriptive bibliometrics in science policy (Noyons 2005; Rafols et al. 2010). The capacity of network visualisation to synthesize complex and multivariate information make them powerful tools for conveying a complex web of relations to diverse stakeholders.

Altmetrics data can be analysed in a variety of different ways and for various goals, as part of supportive materials for case studies. We do not claim that network analysis is necessarily the most appropriate method for analysing almetric data. But as a method to conceptualise, represent and visualise ensembles of interactions among social actors, networks seem particularly apt for analysing engagement among researchers and societal stakeholders. Network visualisations can help in understanding the contexts of researchers through the attributes of the actors (nodes) in their vicinity, and their interactions (links). Engagement of researchers can be hypothesised or inferred from dense networks of interaction with certain stakeholders. In case those stakeholders show traces of behavioural change or innovation, such engagement can be then furthered explored qualitatively as a potential instance of contribution to societal impact. It is only in this hypothetical sense that the contextual mapping provided by altmetric networks informs about potential contribution to impact.

In these analyses, the caveat is that the actual interactions of researchers are not necessarily taking place in social media, nor are they the main channels of influence or engagement. Therefore, we do not assume that the interactions among researchers and stakeholders are the actual engagement. We do assume, however, that researchers with clusters of interactions in social media are likely to be related to engagement – and thus, they provide a lead or hint for case studies to follow.

## 4. Methodological experiment: analysing researchers' networks with Twitter data

We present here a first experiment of the possibilities of Twitter data to capture potential societal engagement of researchers with non-academic communities. The goal is to provide supporting information for case studies on the assessment of societal impact, as part of the palette of methodological tools used within an interaction approach. The idea is that social media can provide added value by showing those non-academic linkages which remain hidden using scientific databases.

### *4.1 Data and methods*

We start by identifying a social media platform in which the researcher under study is active and, at least partially involved with academic contents. In this paper, we use Twitter as the data source for identifying researchers' communities (though other platforms might be more appropriate in other cases). Next, we describe the different steps followed for identifying the communities of each researcher and how we developed and characterised the networks of each of them.

We selected as case studies two researchers of the same research field actively engaged in Twitter, who have very distinct interaction profiles. These case studies were selected to illustrate our argument and they are not random choices. We identified two researchers from the same field and country who would represent two opposite profiles. Based on our previous knowledge of their profile, our experiment confirmed such knowledge. Data for each case study was downloaded fom the public Twitter REST API (https://dev.twitter.com/rest/public). We use the *twitteR* package (Gentry 2015) developed with the statistical programming language R to query the user description of our case studies and download the data.

Next, we identified a list of Twitter users who are followers or followees of the researchers investigated. Then, we identified the community of reciprocal interaction, i.e. the Twitter users who are both followers and followees. This selection is inspired by the notion of 'productive interactions' in the sense that it requires a bidirectional engagement (Spaapen & van Drooge, 2011). However, we do not claim that such link constitutes a productive interaction – it is only a proxy to help contextualise a researcher's network. Moreover, the selection eliminates users like news media, opinion leaders, etc. who are followed but who seldom follow researchers, and which might be considered noise in terms of engagement. This group of users will be what we could define as the interaction community of that given researcher. The number of followers, followees and final interaction community of each case study are shown in Table 2.

**Table 2.** Basic description of the relations of two researchers in Twitter.

| | **Followers** | **Followee** | **Interaction community** |
|---|---|---|---|
| **Researcher 1** | 251 | 81 | 44 |
| **Researcher 2** | 221 | 232 | 110 |

Based on this population of Twitter users, we create a matrix of relations (the data behind the network) between each member with the script *friendsnet.R*. We made the script publicly available at the Github repository (Robinson-Garcia 2017). *friendsnet.R* first searches for the interaction community of a given user and then develops a matrix (network) by querying the list of accounts followed by each user, and then looking for other members of the interaction community in such list. The node of the researcher investigated is removed from the network because she is by definition connected to all other followers/followees.

Finally, we characterise these networks by adding attributes to each node as a way to understand heterogeneity or homogeneity of these communities, and thus the embeddedness (and potential engagement) of the researchers among stakeholders. The specific relevant attributes will depend on the case study. In this paper, we examine two attributes, institutional affiliation and geographical distance, as a means to capture whether the researchers' bridge beyond academia and engage with local policy actors. For geographical location, we consider two categories: whether the follower/followee is local (same country as the researcher) or global (other countries). For institutional affiliation, we classify followers/followee as: academic, private sector, public sector, politicians and NGOs. This information is collected manually based on the information reported in the description of each Twitter account. When this information was not reported, we looked at the user's website when provided. In the case of researcher 2, we were unable to identify the institutional affiliation of six accounts (out of 110).

Figure 1 offers an example of the complete process. From identifying the interaction community (step 1) and developing the network (step 2) to characterising the network by adding attributes to each node based on geographical location (step 3) and institutional affiliation (step 4). Network visualisations are produced using the VOSViewer software developed by van Eck and Waltman (2010) (freely available at http://www.vosviewer.com/).

**Figure 1.** Overview of the methodological approach

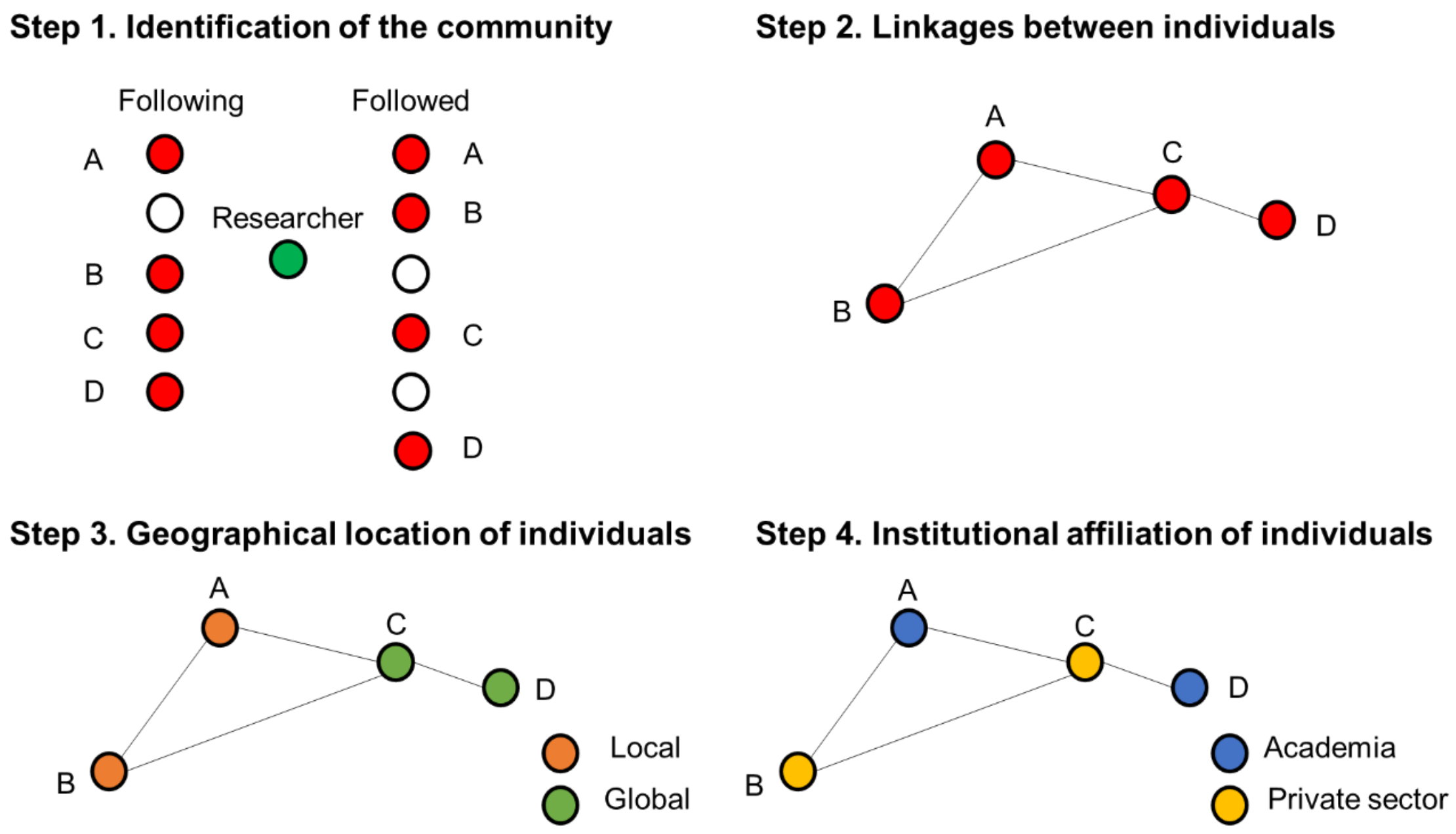

## *4.2. Insights from researchers' networks*

Figure 2 and 3 show the interaction community of two researchers in the same field based on Twitter data. The networks are characterised according to the institutional affiliation of each member in relation to the ego researcher (local or global) and according to their geographical location. Figure 4 displays the share of members of the interaction community by geographical or affiliation attributes.

We observe very different profiles. Researcher 1 exhibits a cohesive interaction community which is mainly international and almost exclusively academic (96%, see Figure 4). Therefore, Twitter data suggests that researcher 1's focus lies in global academic research and that she may not be interested in societal engagement. This conclusion might be a false negative – researcher 1 might be active in societal engagement, but not through Twitter. From the other data sources of the case study, we actually know that researcher 1 is sometimes involved in consultancies and that she is an active blogger. However, her main interests are indeed academic.

Researcher 2 shows an interaction network with two poles: the bottom pole is local community and contains stakeholders from different sectors (45% are non-academic as shown in Figure 4), with presence of public and private sector (mainly consultants). The pole at the top of the figure is more academic. Researcher 2 is strongly embedded in the Dutch science policy community, including public institutions (such as KNAW) and private consultants. The reciprocal following with the policy actors shown by her Twitter interaction community suggests that she is actively engaged in knowledge exchange between academia and policy stakeholders. This information is confirmed from other data sources: Researcher 2 is actively involved in Dutch science policy.

These two cases illustrate that the characterisation of communities based on geographical location and institution affiliation provide a landscape where potential interest groups can be identified. It is thus a helpful starting point for further exploring interactions between academics and non-academics in a case study. Since non-academics can be active players in the Twitter network, this approach does not conceive the relation between academia and stakeholders as unidirectional or hierarchical. It adopts an open approach which goes in line with the conceptual notion of open science and of academia as an institution which can be entangled (or disentangled, as in Researcher 1's case) with various or specific societal stakeholders --potentially responding or not to the societal demands of specific stakeholders' (Hess 2016).

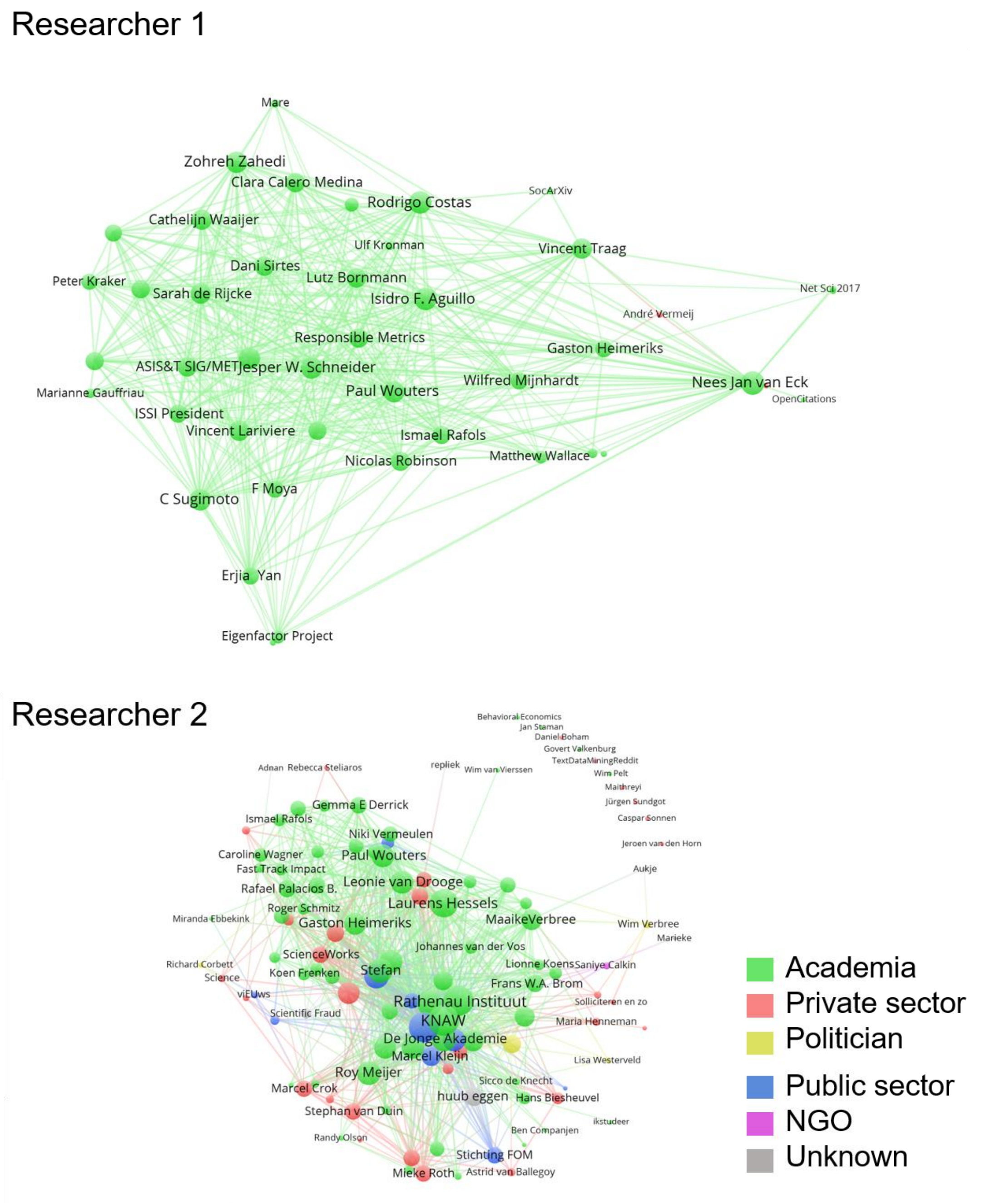

**Figure 2.** Interaction communities of two researchers (names not shown) from the same field using Twitter data of their reciprocal follower relations, characterised by the institutional affiliation of nodes

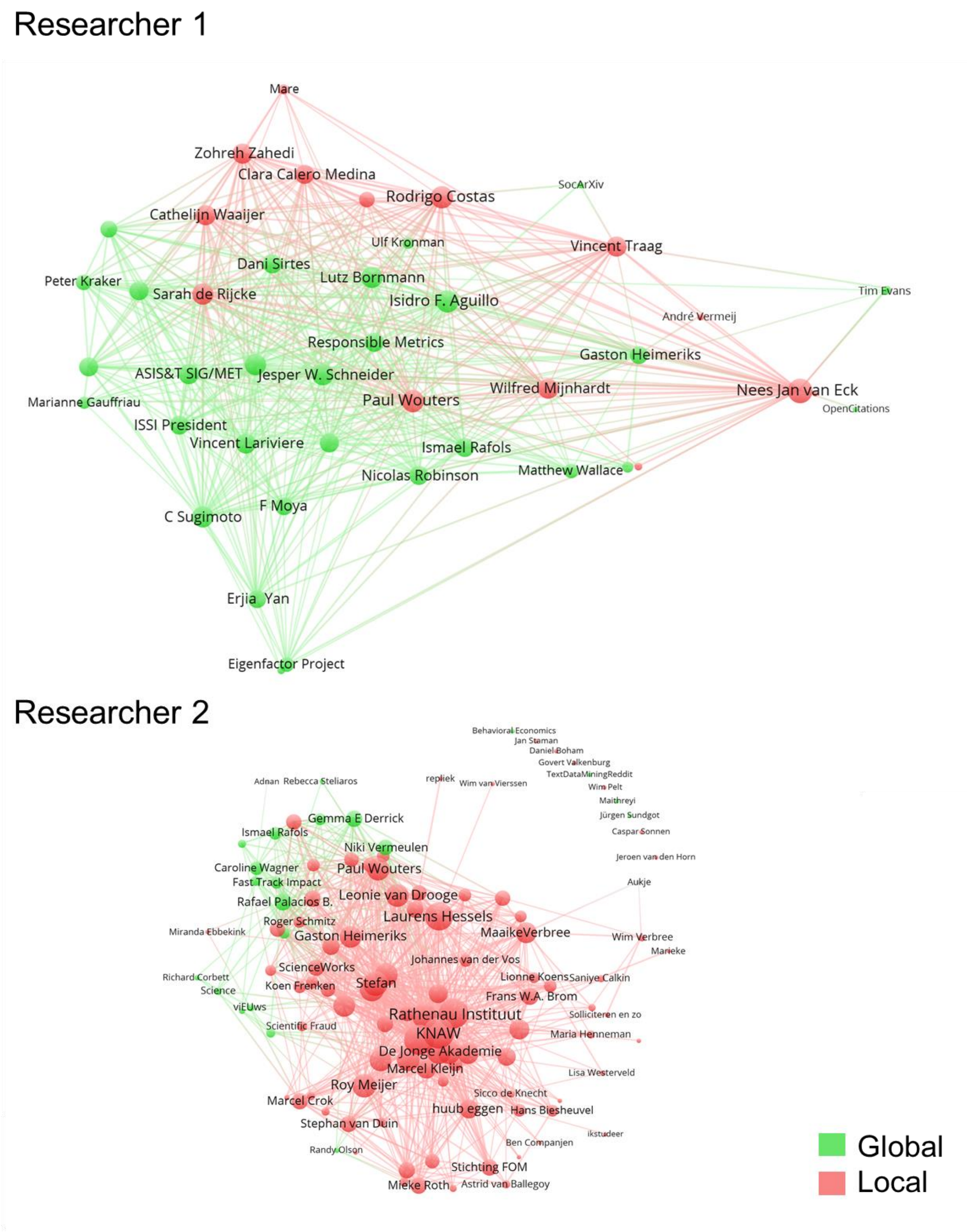

**Figure 3.** Interaction communities of two researchers (names not shown) from the same field using Twitter data of their reciprocal relations, characterised by their geographical location with respect to the researcher (where local means from the same country, i.e. the Netherlands).

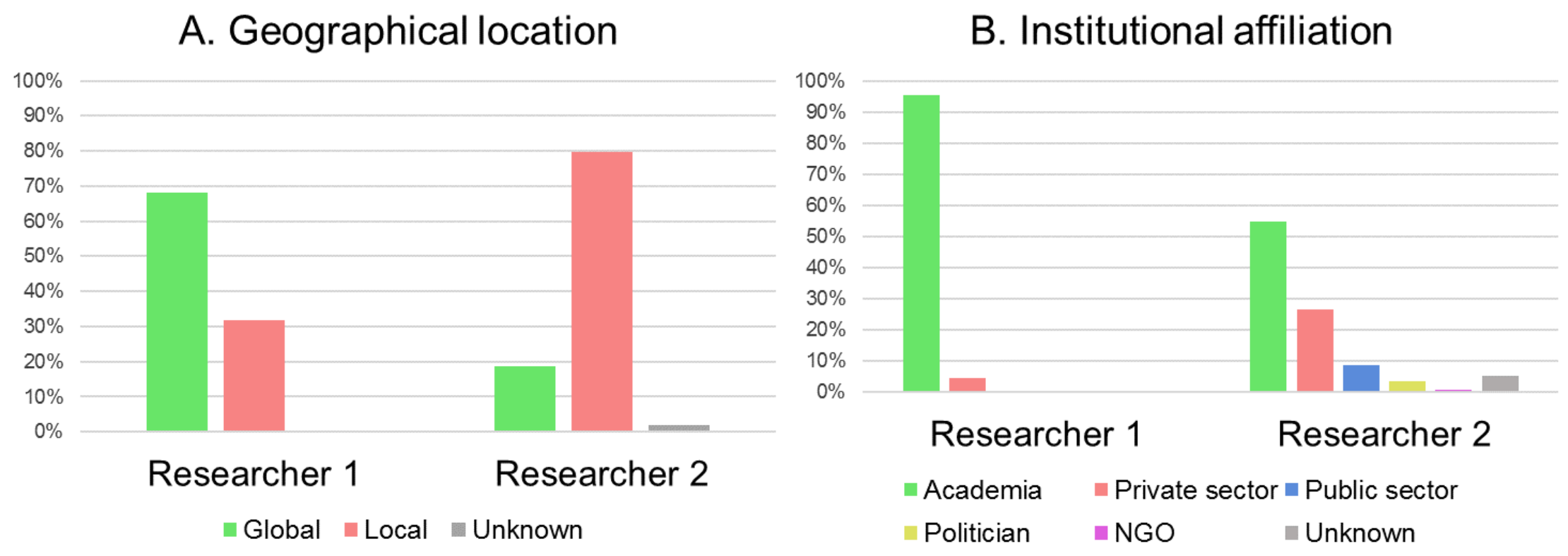

**Figure 4.** Share of members of the interaction community using Twitter data characterised by their geographical location (A) and their institutional affiliation (B).

This type of analyses, which consider researchers' context and their capacity to bridge between communities, can serve as a means to identify researchers with a policy or orientation to local stakeholders. While the aim in this example was to capture interactions between scholars and non-academics, the method has the potential to be applied in other contexts. Generally, these descriptions can help characterise the nature and outreach of a researchers' activity through the analysis of the communities in which they are embedded. For example, they can also be useful for the exploration of informal communication in invisible colleges within purely academic networks (Crane, 1972), or among scientific experts involved in international policy (Haas, 1992).

This method is subject to important limitations. First, the attributes of the nodes of the network were manually collected and coded (what makes scaling up difficult). The identification of subjects is problematic as users may be institutions, anonymous or individuals. Even when they are individuals, it was not straightforward to identify their affiliation or profile, nor to characterise them. Recently, some technical advances have been reported in this respect (Alperín, 2015b). Second, since many researchers do not use Twitter, it is not a general tool that can be universally employed, as bibliometric indicators, what is more, when used it is also employed in many cases for personal matters. In some case studies, other social media rather than Twitter might be more relevant. In some other cases, social media will yield no signal and qualitative methods will be needed. Hence, the method is aimed to be used as an exploratory tool in case studies with some previous understanding of the contexts. Third, in social media there are no clear boundaries between an academics' professional and personal activities and interactions. Social media mix both aspects, which may be both a strength and a weakness of the approach, but makes ambiguous the meaning of the network links.

# 5. Conclusions

In this article, we have proposed a methodological framework to analyse altmetric data in order to inform assessment on the societal impact of research. In contrast to current altmetric analyses which are extrapolated from citation frameworks (Haustein et al. 2015), this methodology is based on interaction approaches conceptually grounded in impact evaluation scholarship such as SIAMPI or ASIRPA (Spaapen & van Drooge, 2011; Joly et al., 2015). The emphasis of the evaluation is on learning and improving societal impact. It focuses on understanding and mapping, through case studies, the contexts in which engagement among researchers and stakeholders might take place – assuming that engagement is likely to lead to impact. In doing so, it departs from universal or generalizable methods with a focus on accountability, for which altmetric data are not sufficiently robust and could be counterproductive.

An analysis of the policy context suggests that there is an important demand for altmetrics to develop universal and scalable methodologies for assessing societal impact of research. However, these methodologies must be grounded in an understanding of the processes leading to societal impact (Molas-Gallart & Davies 2006; Joly et al. 2015). This involves developing theories on social media use of scholarly information (Haustein et al., 2015), as well as theories on how societal impact occurs (Joly et al. 2015) and about the policy context of evaluation (Wouters, 2014). Otherwise, these new techniques and approaches could inherit and suffer from the same (or possibly worse) limitations and biases of the traditional methodologies applied to identify scientific impact (Hicks et al. 2015). We believe that this may be the case for many altmetric indicators, as they are currently conceived and used. Although their application may have seemed promising, they are based on the same theoretical premises than citation indicators, thus leading to a narrow understanding of the researchers' societal impact.

However, social media contain potentially rich information to capture evidence of 'broader impact'. Social media are powerful in that their heterogeneity greatly broadens the scope of science-society interactions mapped, in a way that no other source does nowadays. It is in these connections – in the processes of communication and interaction-- where their interest for societal impact assessment lies.

We have argued that, for altmetric data to be useful for societal impact assessment, a shift on evaluation goals and methodologies is also necessary (Spaapen 2015). Methodologies that have so far been largely qualitative can provide a useful conceptual framework for using social media data as input to societal impact assessment. This family of approaches acknowledges that societal impact cannot be measured in a standardised and universally applicable way. Instead they suggest to carry out case studies focusing on a contextualised mapping of the interactions among researchers and stakeholders which may lead to societal impact (Joly et al., 2015). Within these type of approaches, altmetric data could be used as suggestions or hints for further research on potential societal impact. It might also be useful as triangulation of evidence with other data sources or as an illustration of evidence obtained from more reliable methods (e.g. ethnographies).

One of the potential ways to use altmetric data is to map the networks of interaction among researchers and stakeholders. As an example, here we have used Twitter data, looking at reciprocal follower-followee interactions –which is a potential trace of two-way communication. Using these links, we have built the Twitter interaction networks of two researchers. These networks serve as a first approximation towards understanding the contexts of researchers in social media. While the networks do not directly reflect societal engagement, they suggest areas of potential productive interactions. They can also be helpful to explore or illustrate the mechanisms and processes that enable societal impact. This network approach facilitates to shift from an assessment of attribution towards an assessment of contribution to impact, as proposed by Spaapen and van Drooge (2011). The approach shown is still limited and could be further refined by incorporating other information such as the number of tweets researchers and their community exchange with each other, searching for similarity patterns on the type of information they share and analysing the tweet contents. The two examples illustrate the potential of the method to suggest the researchers' contributions in terms of the types of communities they engage with – which are very likely to be related to their type of scientific and/or societal impact.

Further research is needed, given the very uneven participation and coverage of researchers in Twitter. Questions to address in the near future concern the use of these methods for supporting case studies within larger methodologies such as PIPA (Douthwaite et al., 2007), and the perceptions of validity and usefulness by policy-makers pressured towards accountability approaches. Further technical questions for the characterisation of the altmetric networks involve the characterisation of Twitter accounts, currently based on manual coding; and whether a distinction between professional and personal communications would be feasible or desirable.

We believe that these mapping and contextual approaches to altmetrics can be particularly relevant for exploring the impacts (both academic and societal) of the social sciences and the humanities. These are scholarly areas which are very badly covered by scientific databases and societal impact indicators, because non-standard publications (Díaz-Faes et al. 2016) and informal interactions regarding socioeconomic and cultural issues are more common than in other fields (Olmos et al. 2013). Another issue of special interest is the use of these approaches to explore the engagement of researchers with local or global peers or stakeholders, an issue that is hotly debated in countries in the 'scientific peripheries' with pressure to publish internationally (Piñeiro & Hicks 2015; Chavarro et al. 2017).

# Acknowledgments

Preliminary results of this paper were presented at the STI Conference 2016 and Open Evaluation Conference. We thank Rodrigo Costas, Jordi Molas-Gallart and Irene Ramos-Vielba for helpful comments and discussions. We also thank two anonymous reviewers for their helpful comments and suggestions. Nicolas Robinson-Garcia is currently supported by a Juan de la Cierva-Formación postdoctoral grant from the Spanish Ministry of Economy and Competitiveness.

# References

Alperin, J. P. (2015a). 'Geographic variation in social media metrics: an analysis of Latin American journal articles', *Aslib Journal of Information Management*, 67/3: 289–304.

Alperin, J. P. (2015b). 'Moving beyond counts: A method for surveying Twitter users', <http://altmetrics.org/altmetrics15/alperin/> accessed 21 Sep, 2016.

Bonney, R., Cooper, C. B., Dickinson, J., Kelling, S., Phillips, T., Rosenberg, K. V. and Shirk, J. (2009). 'Citizen Science: A Developing Tool for Expanding Science Knowledge and Scientific Literacy', *BioScience*, 59/11: 977–84.

Bornmann, L. (2014). 'Do altmetrics point to the broader impact of research? An overview of benefits and disadvantages of altmetrics', *Journal of Informetrics*, 8/4: 895–903.

Cabezas-Clavijo, Á. and Torres-Salinas, D. (2010). 'Indicadores de uso y participación en las revistas científicas 2.0: el caso de PLoS One', *El profesional de la información*, 19/4: 431–434.

Callon, M. (1986). 'The sociology of an actor-network: The case of the electric vehicle', *Mapping the dynamics of science and technology*, pp. 19-34. Palgrave Macmillan UK.

Chavarro, D. A., Tang, P. and Rafols, I. (2016). 'Why researchers publish in non-mainstream journals: Training, knowledge bridging, and gap filling'. *SPRU Working Paper Series*, 2016-22. <https://papers.ssrn.com/sol3/papers.cfm?abstract_id=2887274> accessed 10 Jan, 2017.

Costas, R., van Honk, J., Zahedi, Z. and Calero-Medina, C. (2016) 'Discussing practical applications for altmetrics: social media profiles for African, European and North American publications'. *Presentation at the Conference 3:AM, Bucharest, September 2016.* <https://doi.org/10.6084/m9.figshare.3980145.v1> accessed 10 Jan, 2017.

Costas, R, Zahedi, Z. and Wouters, P. (2015). 'The thematic orientation of publications mentioned on social media: Large-scale disciplinary comparison of social media metrics with citations', *Aslib Journal of Information Management*, 67/3: 260–288.

Crane, D. (1972). *Invisible colleges: Diffusion of knowledge in scientific communities*. Chicago: University of Chicago Press.

*Crowdsourcing and Citizen Science Act of 2015* (2015). Vol. MDM15G24. < https://www.congress.gov/bill/114th-congress/senate-bill/2113> accessed on 3 Oct, 2016.

De Jong, S. P. L., van Arensbergen, P., Daemen, F., van der Meulen, B. and van den Besselaar, P. (2011). 'Evaluation of research in context: an approach and two cases', *Research Evaluation*, 20/1: 61–72.

Díaz-Faes, A. A., Bordons, M. and van Leeuwen, T. N. (2016). 'Integrating metrics to measure research performance in social sciences and humanities: The case of the Spanish CSIC', *Research Evaluation*, 25/4: 451-460.

Donovan, C. and Butler, L. (2007). 'Testing novel quantitative indicators of research "quality", esteem and "user engagement": an economics pilot study', *Research Evaluation*, 16/4: 231–42.

Douthwaite, B., Alvarez, S., Cook, S., Davies, R., George, P., Howell, J. et al. (2007). 'Participatory impact pathways analysis: a practical application of program theory in research-for-development', *The Canadian Journal of Program Evaluation*, 22/2: 127.

Freeman, C. (1991). 'Networks of innovators: a synthesis of research issues'. *Research policy*, 20/5: 499-514.

Frodeman, R. and Parker, J. (2009). 'Intellectual merit and broader impact: The National Science Foundation's broader impacts criterion and the question of peer review', *Social Epistemology*, 23/3–4: 337–345.

Gentry, J. (2015). twitteR (v. 1.1.8). R CRAN v. 3.3.0. Available at <https://github.com/geoffjentry/twitteR>.

Geuna, A. and Muscio, A. (2009). ‘The Governance of University Knowledge Transfer: A Critical Review of the Literature’, *Minerva,* 47/1: 93–114. DOI: 10.1007/s11024-009-9118-2

Grande, D., Gollust, S. E., Pany, M., Seymour, J., Goss, A., Kilaru, A. and Meisel, Z. (2014). ‘Translating Research For Health Policy: Researchers’ Perceptions And Use Of Social Media’, *Health Affairs,* 33/7: 1278–85.

Haas, P. M. (1989). ‘Do regimes matter? Epistemic communities and Mediterranean pollution control’, *International organization*, 43/3: 377-403.

Haustein, S., Bowman, T. D. and Costas, R. (2015). ‘Interpreting “altmetrics”: viewing acts on social media through the lens of citation and social theories’. Sugimoto, C. R. (ed.), *Theories of Informetrics and Scholarly Communication: A Festschrift in honor of Blaise Cronin*, pp. 372-406. De Gruyter.

Hazelkorn, E. (2011). ‘Rankings and the Reshaping of Higher Education: The Battle for World-Class Excellence.’, *Palgrave Macmillan*.

Hess, D. J. (2016). *Undone Science: Social Movements, Mobilized Publics, and Industrial Transitions*. MIT Press.

Holbrook, J. B. (2012). ‘Re-assessing the science-society relation: The case of the US National Science Foundation’s broader impacts merit review criterion (1997-2011)’, *Techonology in Society,* 27/4: 437–451.

Hsu, C. and Park, H. W. (2011). ‘Sociology of Hyperlink Networks of Web 1.0, Web 2.0, and Twitter: A Case Study of South Korea’, *Social Science Computer Review,* 29/3: 354–68.

Ison, R. and Russell, D. (2007). *Agricultural extension and rural development: breaking out of knowledge transfer traditions*. Cambridge University Press.

Joly, P. B., Gaunand, A., Colinet, L., Larédo, P., Lemarié, S. and Matt, M. (2015). ‘ASIRPA: A comprehensive theory-based approach to assessing the societal impacts of a research organization’. *Research Evaluation*, 24/4: 440-453.

Ke, Q., Ahn, Y.-Y., & Sugimoto, C.R. (2017). A systematic identification and analysis of scientists on Twitter. Plos One, 12(4), e0175368.

Kok, M. O. and Schuit, A. J. (2012). ‘Contribution mapping: a method for mapping the contribution of research to enhance its impact’. *Health Research Policy and Systems,* 10/1: 21.

Meijer, I., Mejlgaard, N., Lindner, R., Woolley, R., Rafols, I., Griesler, E., Wroblewski, A., et al. (2016). ‘Monitoring the Evolution and Benefits of Responsible Research and Innovation (MoRRI) – a preliminary framework for RRI dimensions & indicators’. Presented at the OECD Blue Sky Forum 2016, Ghent.

Moed, H. F. (2015). ‘Altmetrics as traces of the computerization of the research process’. Sugimoto, C. R. (ed.), *Theories of Informetrics and Scholarly Communication: A Festschrift in honor of Blaise Cronin*, pp. 360-371. De Gruyter.

Moed, H. F. and Halevi, G. (2015). ‘The Multidimensional Assessment of Scholarly Research Impact’, *Journal of the Association for Information Science and Technology*, 66/10: 1988–2002.

Moedas, C. (2015). *Open Innovation, Open Science, Open to the World*.

Molas-Gallart, J. and Castro-Martínez, E. (2007). ‘Ambiguity and conflict in the development of “Third Mission” indicators’, *Research Evaluation,* 16/4: 321–30.

Molas-Gallart, J. and Davies, A. (2006). ‘Toward theory-led evaluation the experience of European science, technology, and Innovation policies’. *American Journal of Evaluation*, 27/1: 64-82.

Molas-Gallart, J., D’Este, P., Llopis, O. and Rafols, I. (2015). ‘Towards an alternative framework for the evaluation of translational research initiatives’, *Research Evaluation*, 25/3: 235-243.

Molas-Gallart, J., Salter, A., Patel, P., Scott, A. and Duran, X. (2002). Measuring third stream activities. *Final report to the Russell Group of Universities. Brighton: SPRU, University of Sussex*. <http://s3.amazonaws.com/academia.edu.documents/3460866/russell_report_thirdStream.pdf?AWSAccessKeyId=AKIAIWOWYYGZ2Y53UL3A&Expires=1489080056&Signature=%2FiL4tymktBQ6pqf6b0N4eHwMlvs%3D&response-content-disposition=inline%3B%20filename%3DMeasuring_third_stream_activities.pdf> accessed on 10 Feb, 2017.

Noyons, C. M. (2005). ‘Science Maps Within a Science Policy Context’. Moed H. F., Glänzel W., & Schmoch U. (eds) *Handbook of Quantitative Science and Technology Research*, pp. 237–55. Springer Netherlands.

Nowotny, H., Scott, P., & Gibbons, M. (2001). *Re-thinking science: knowledge and the public in an age of uncertainty*. Wiley.

Owen, R., Macnaghten, P., & Stilgoe, J. (2012). ‘Responsible research and innovation: From science in society to science for society, with society’, *Science and Public Policy*, 39/6: 751–60.

Olmos-Peñuela, J., Molas-Gallart, J. and Castro-Martínez, E. (2013). 'Informal collaborations between social sciences and humanities researchers and non-academic partners', *Science and Public Policy*, 41: 493-506.

Piñeiro, C. L. and Hicks, D. (2015). 'Reception of Spanish sociology by domestic and foreign audiences differs and has consequences for evaluation'. *Research Evaluation*, 24/1: 78-89.

Piwowar, H. (2013). 'Altmetrics: Value all research products', *Nature*, 493/7431: 159–159.

Priem, J. (2010). 'I like the term #articlelevelmetrics, but it fails to imply *diversity* of measures. Lately, I'm liking #altmetrics.'

Priem, J., Piwowar, H. A. and Hemminger, B. M. (2012). 'Altmetrics in the wild: Using social media to explore scholarly impact', *arXiv:1203.4745 [cs]*.

Priem, J., Taraborelli, P., Groth, C., & Neylon, C. (2010). 'altmetrics: a manifesto – altmetrics.org'. <http://altmetrics.org/manifesto/> accessed 13 Feb, 2014.

Procter, R., Williams, R., Stewart, J., Poschen, M., Snee, H., Voss, A. and Asgari-Targhi, M. (2010). 'Adoption and use of Web 2.0 in scholarly communications', *Philosophical Transactions of the Royal Society of London A: Mathematical, Physical and Engineering Sciences*, 368/1926: 4039–56.

Rafols, I., Ciarli, T., van Zwanenberg, P. and Stirling, A. (2012). 'Towards indicators for "opening up" science and technology policy'. *Proceedings of 17th International Conference on Science and Technology Indicators*, Vol. 2, pp. 663–74.

Rafols, I., Porter, A. L. and Leydesdorff, L. (2010). 'Science overlay maps: A new tool for research policy and library management', *Journal of the American Society for information Science and Technology*, 61/9: 1871–1887.

Reale, E., Primeri, E., Flecha, R., Soler, M., Oliver, E., Puigvert, L., UAB, T. S., et al. (2014). *Report 1. State of the art in the scientific, policy and social impact of SSH research and its evaluation*. <http://impact-ev.eu/wp-content/uploads/2015/08/D1.1-Report-1.-State-of-the-art-on-scientific-policy-and-social-impact-of-SSH-research-and-its-evaluation.pdf> accessed 15 Jan, 2016.

REF (2011). Assessment framework and guidance on submissions ( updated addendum published in January 2012). <http://www.ref.ac.uk/media/ref/content/pub/assessmentframeworkandguidanceonsubmissions/GOS%20including%20addendum.pdf> accessed 13 Jan, 2017.

Robinson-García, N., Torres-Salinas, D., Zahedi, Z. and Costas, R. (2014). 'New data, new possibilities: exploring the insides of Altmetric. com', *El profesional de la información*, 23/4: 359–366.

Robinson-Garcia, N., Trivedi, R., Costas, R., Isett, K., Melkers, J., & Hicks, D. (forthcoming). The unbearable emptiness of tweeting – about journal articles. Submitted.

Robinson-Garcia, N. (2017). 'twitter-networks'. *GitHub repository*. <https://github.com/elrobin/twitter-networks> accessed 10 Feb, 2017.

Ronald, R. and Fred, Y. Y. (2013). 'A multi-metric approach for research evaluation', *Chinese Science Bulletin*, 58/26: 3288–90.

Samuel, G. N. and Derrick, G. E. (2015). 'Societal impact evaluation: Exploring evaluator perceptions of the characterization of impact under the REF2014', *Research Evaluation*, 24/3: 229–41.

Schroeder, R. (2007). 'e-Research Infrastructures and Open Science: Towards a New System of Knowledge Production?', *Prometheus*, 25/1: 1–17.

Spaapen, J. B. (2015). 'A new evaluation culture is inevitable', *Organic Farming*, 1/1: 36-37.

Spaapen, J. and van Drooge, L. (2011). 'Introducing "productive interactions" in social impact assessment', *Research Evaluation*, 20/3: 211–8.

Sugimoto, C. R., Work, S., Larivière, V. and Haustein, S. (2016). 'Scholarly use of social media and altmetrics: a review of the literature', *arXiv preprint arXiv:1608.08112*.

Tenopir, C., Volentine, R. and King, D. W. (2013). 'Social media and scholarly reading', *Online Information Review*, 37/2: 193–216.

Thelwall, M. (2012). 'A history of webometrics', *Bulletin of the American Society for Information Science and Technology*, 38/6: 18–23.

Van Eck, N.J., & Waltman, L. (2010). Software survey: VOSviewer, a computer program for bibliometric mapping. *Scientometrics*, 84/2: 523-538.

Van Leeuwen, T. N. and Moed, H. F. (2012). 'Funding decisions, peer review, and scientific excellence in physical sciences, chemistry, and geosciences', *Research Evaluation*, 21/3: 189–98.

Wallace, M. L. and Rafols, I. (2015). 'Research Portfolio Analysis in Science Policy: Moving from Financial Returns to Societal Benefits', *Minerva*, 53/2: 89–115.

Wilsdon, J., & al. (2015). *The Metric Tide: Report of the Independent Review of the Role of Metrics in Research Assessment and Management.* < http:// 10.13140/RG.2.1.4929.1363> accessed on 26 Sep, 2016.

Wouters, P. (2014). ‘The citation: From culture to infrastructure’. Cronin, B., and Sugimoto, C. (eds) *Beyond bibliometrics: harnessing multidimensional indicators of scholarly impact*, pp. 47-66. MIT Press.

Zahedi, Z., Fenner, M. and Costas, R. (2014). ‘How consistent are altmetrics providers? Study of 1000 PLOS ONE publications using the PLOS ALM, Mendeley and Altmetric. com APIs’. Altmetrics 14. Workshop at the Web Science Conference, Bloomington, USA.